\documentclass[journal]{IEEEtran}
\usepackage{amsfonts}

\usepackage{ifpdf}
\usepackage{cite}
\usepackage{url}

\ifCLASSINFOpdf
   \usepackage[pdftex]{graphicx}
\else
\fi

\usepackage[cmex10]{amsmath}

\usepackage{array}
\usepackage{mdwmath}
\usepackage{mdwtab}
\usepackage{eqparbox}
\usepackage[tight,footnotesize]{subfigure}
\usepackage{algorithm}
\usepackage{amsmath}
\usepackage{epsfig}
\usepackage{ae}
\usepackage{amssymb}
\usepackage{times}
\usepackage{color}
\usepackage{verbatim}
\usepackage{mathrsfs}
\usepackage{color}
\usepackage{epstopdf}
\usepackage{longtable}
\usepackage{float}
\usepackage{array, booktabs}
\usepackage{algorithm}
\usepackage{algpseudocode}
\usepackage{amsmath}
\usepackage{comment}

\def\x{\boldsymbol x}

\def\0{\boldsymbol 0}
\def\1{\boldsymbol 1}

\begin{document}

\title{ \huge Accurate RSS-Based Localization Using an Opposition-Based Learning Simulated Annealing Algorithm

\thanks{This work was supported in part by the National Natural Science Foundation of China (No. 61671260), in part by the Research Startup Fund of Ningbo University of Technology (No. 2040011540012), in part by the Scientific Research Fund of Zhejiang Provincial Education Department (No. Y201839519), in part by the Ningbo Natural Science Foundation (No. 2019A610087).}
\thanks{The authors are with the School of Cyber Science and Engineering, Ningbo University of Technology, Ningbo 315211, China (e-mail: weizhongdingnubt@163.com, csm20130504@163.com, shudi.bao@nbut.edu.cn, nbchen75@sina.com, sunjie@nbut.edu.cn.)(Corresponding author: Shengming Chang) }}%

\author{Weizhong Ding, \IEEEmembership{Student Member, IEEE}, Shengming Chang, Shudi Bao,  Meng Chen, and Jie Sun}

\maketitle
\date{}
\begin{abstract}
Wireless sensor networks require accurate target localization, often achieved through received signal strength (RSS) localization estimation based on maximum likelihood (ML). However, ML-based algorithms can suffer from issues such as low diversity, slow convergence, and local optima, which can significantly affect localization performance. In this paper, we propose a novel localization algorithm that combines opposition-based learning (OBL) and simulated annealing algorithm (SAA) to address these challenges. The algorithm begins by generating an initial solution randomly, which serves as the starting point for the SAA. Subsequently, OBL is employed to generate an opposing initial solution, effectively providing an alternative initial solution. The SAA is then executed independently on both the original and opposing initial solutions, optimizing each towards a potential optimal solution. The final solution is selected as the more effective of the two outcomes from the SAA, thereby reducing the likelihood of the algorithm becoming trapped in local optima. Simulation results indicate that the proposed algorithm consistently outperforms existing algorithms in terms of localization accuracy, demonstrating the effectiveness of our approach.

\end{abstract}
\begin{IEEEkeywords}
Localization, wireless sensor networks, received signal strength, simulated annealing algorithm, opposition-based learning.
\end{IEEEkeywords}

\section{Introduction}\label{s1}
Target localization in wireless sensor networks (WSNs) has gained significant attention due to its practical importance and diverse applications across various domains, such as military, health, environment, home, business, regional monitoring, health care monitoring, environmental/earth sensing, air pollution monitoring, forest fire detection, landslide detection, water quality monitoring, and industrial monitoring \cite{WSNs1,WSNs2,WSNs3,WSNs4}. 
The most commonly utilized parameters for localization include received signal strength (RSS), angle of arrival (AoA), time of arrival (ToA), and time difference of arrival (TDoA).  Among these, RSS is particularly popular due to its low cost. In WSNs, sensor nodes can be classified into two types: anchor nodes and target nodes. Anchor nodes have known locations, while target nodes have unknown locations. Typically, anchor nodes transmit radio signals to target nodes, and the signal strengths received by the target nodes are measured as RSS values. Based on the anchor node locations and the RSS values, target localization can be achieved. However, RSS values are imprecise due to measurement noise, which can be caused by environmental disturbances and the inherent variability in wireless signal propagation. To model this optimization problem, a nonlinear, non-convex \cite{non-convex} maximum likelihood (ML) estimator is often employed.

RSS based localization is indeed an optimization process that seeks the optimal solution to a localization problem using a specific algorithm. Recent studies have employed convex optimization techniques, such as second-order cone programming  (SOCP) \cite{SOCP} and semi-definite programming (SDP) \cite{SR-SDP}, to relax the non-convex estimator to  convex ones. Despite their efficiency, the algorithms still faces challenges in real-time applications due to their complexity. A. Beck et al. proposed a fast and low-complexity localization algorithm in \cite{GTRS}. They developed a squared-range-based weighted least squares (SR-WLS) estimator, which can be derived from a generalized trust region subproblem (GTRS). They efficiently solved the GTRS using a bisection procedure. Although the classical optimization techniques can find the optima of problems with specific mathematical features, such as continuity, differentiability, convexity, and unimodality. However, many real-world optimization problems lack these features. To address such cases, various metaheuristic optimization techniques have emerged in recent years. 

Recently, some researchers have employed heuristic intelligent optimization algorithms to solve the optimization problem of target localization. LAC. Najarro et al. presented an exact algorithm \cite{DEOR} based on differential evolution with opposition and redirection (DEOR). They utilizd differential evolution to solve the ML estimation problem and applied opposition-based learning and redirection to enhance accuracy. In their latest work \cite{DEORfast}, they reduced the algorithm complexity by using opposition-based learning (OBL) only for generating the initial population. We refer to this algorithm as ``DEOR-fast'' in the rest of this paper. Although the above algorithms can find the global optima for many localization problems. However, they often require modifications to enhance their performance, especially when the measurement noise intensity is low. This observation aligns with the No Free Lunch (NFL) theorem \cite{NFL}, which posits that no single algorithm can be optimal for all optimization problems. This theorem has spurred the development of optimization algorithms, motivating researchers to modify existing algorithms and improve their performance. Inspired by these two studies, we applied OBL to the simulated annealing algorithm (SAA), which remarkably increased the probability of finding the global optimal solution.

In this paper, we enrich prior work by integrating the SAA into the localization algorithm. Although SAA is an effective local search strategy, it may fall into local optima in complex search spaces.  To overcome this potential pitfall, we introduce the OBL technique into our algorithm. OBL generates an opposing initial solution to the original one, thereby enhancing the diversity of the search space \cite{OBLglobaloptim}. Continuing the process, our algorithm utilizes the SAA to independently execute searches starting from both the original initial solution and its opposing counterpart. Then, the algorithm selects the superior solution between the two final outcomes. This allows the solution to escape from its local optima of attraction with a certain probability, leading to a higher likelihood of converging to the global optimum.

This paper presents significant contributions to the field of target localization in wireless sensor networks. Our primary innovation lies in the development of a novel strategy that combines the SAA and OBL to effectively refine a randomly generated initial location of the target node within the solution space. The introduction of an opposing solution, generated by leveraging OBL, considerably enhances the search capability for the global optimum. Our simulation results underscore the efficacy of the proposed strategy in achieving high localization accuracy, making it a compelling approach for applications requiring precise target localization.


\section{RSS Measurement Model and Problem Formulation}\label{s2}
We consider a two-dimensional WSN with $N$ anchor nodes that have known locations $\boldsymbol{a}_i=[a_{i1}, a_{i2}]^\mathrm{T}, i=1, 2, \cdots, N$, and one target node that has an unknown location $\boldsymbol{x}=[x_1,x_2]^\mathrm{T}$. We model the RSS measurements, $P_i$, between the $i$-th anchor node and the target node as 
\begin{align}\label{e1}
P_i=P_0-10\gamma {\log _{10}}\frac{\left\| \boldsymbol{x}-{\boldsymbol{a}_i} \right\|}{d_0} +n_i, \quad  i=1, 2, \cdots, N,
\end{align}
where $P_0$ is the reference received signal power at the reference distance $d_0$ and $\gamma$ is the path loss exponent. $n_i$ is the measurement noise of $P_i$, which we assume it to be a zero-mean Gaussian random variable, a presumption consistent with common practice.

The localization problem is to estimate the target node’s location, $\boldsymbol{x}$. We formulate the ML estimator of $\boldsymbol{x}$ as 
\begin{align}\label{e2}
\underset{\boldsymbol{x}}{\mathop{\min}}\,\sum\limits_{i=1}^{N}{\frac{{\left({P_i}-{P_0}+10\gamma {{\log }_{10}}\frac{\left\| \boldsymbol{x}-{\boldsymbol{a}_i} \right\|}{d_0}\right)}^2}{{\sigma _{i}}^2}}, 
\end{align}
where $\sigma _{i}$ is the standard deviation of noise $n_i$.

\section{Proposed Method Derivation}\label{s3}
This section introduces an accurate localization algorithm based on the SAA and the OBL. In the proposed algorithm, we start with an original initial solution, which is generated randomly in the solution space. An opposing initial solution is then created as an alternative starting point using OBL. As the algorithm progresses, we generate original intermediate solutions and opposing intermediate solutions from the original and opposing initial solutions, respectively. Finally, the original final solution and opposing final solution are the outcomes at the end of the optimization process, which are derived from the original and opposing initial solutions, respectively.

We randomly obtain the original initial solution, $\hat{\boldsymbol{x}}_0$, of the SAA process as 
\begin{align}
\hat{\boldsymbol{x}}_0 = \boldsymbol{r} \odot (\x_{\text{max}} - \x_{\text{min}}) + \x_{\text{min}},
\end{align}
where $\boldsymbol{r}=[r_1,r_2]^\mathrm{T}$, $r_1$ and $r_2$ are two random variables, each uniformly distributed in the interval $[0, 1]$. Symbol $\odot$ denotes the Hadamard product. $\x_{\text{min}}$ and $\x_{\text{max}}$ are the boundaries of the solution space.

\begin{figure}
\centering
\includegraphics[width=1\linewidth]{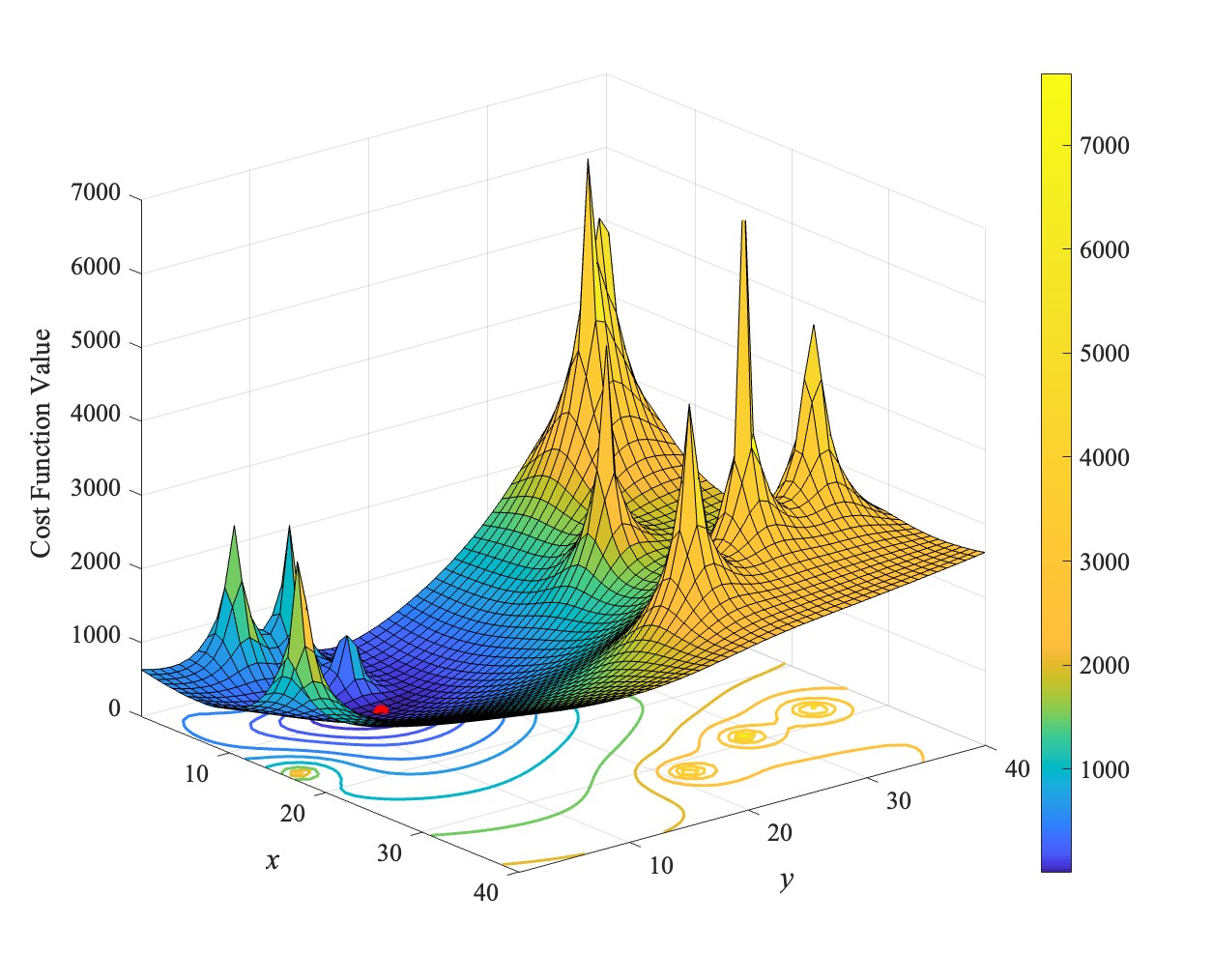}
\caption{The cost function of the ML estimator.}\label{map3D}
\end{figure}

While the SAA demonstrates effective search performance in numerous complex optimization problems, it faces challenges when addressing the cost function of the RSS localization problem, as depicted in Figure.\,\ref{map3D}, where the target node is located at the red node’s location. This function is a complex multi-peak function riddled with numerous local optima. In such a scenario, SAA might become trapped in local optima. This not only impacts the global search capability of the algorithm but could also significantly affect the localization accuracy.

To mitigate this issue, we introduce the OBL technique. By generating an opposing initial solution, $\overline{\boldsymbol{x}}_0$, we anticipate enhancing the diversity of the search space, thereby increasing the likelihood of locating the global optimum. More specifically, $\overline{\boldsymbol{x}}_0$ is generated as follows,
\begin{align}
\overline{\boldsymbol{x}}_0=\x_{\text{max}}+\x_{\text{min}}-\hat{\boldsymbol{x}}_0.
\end{align}

\begin{figure}
\centering
\includegraphics[width=1\linewidth]{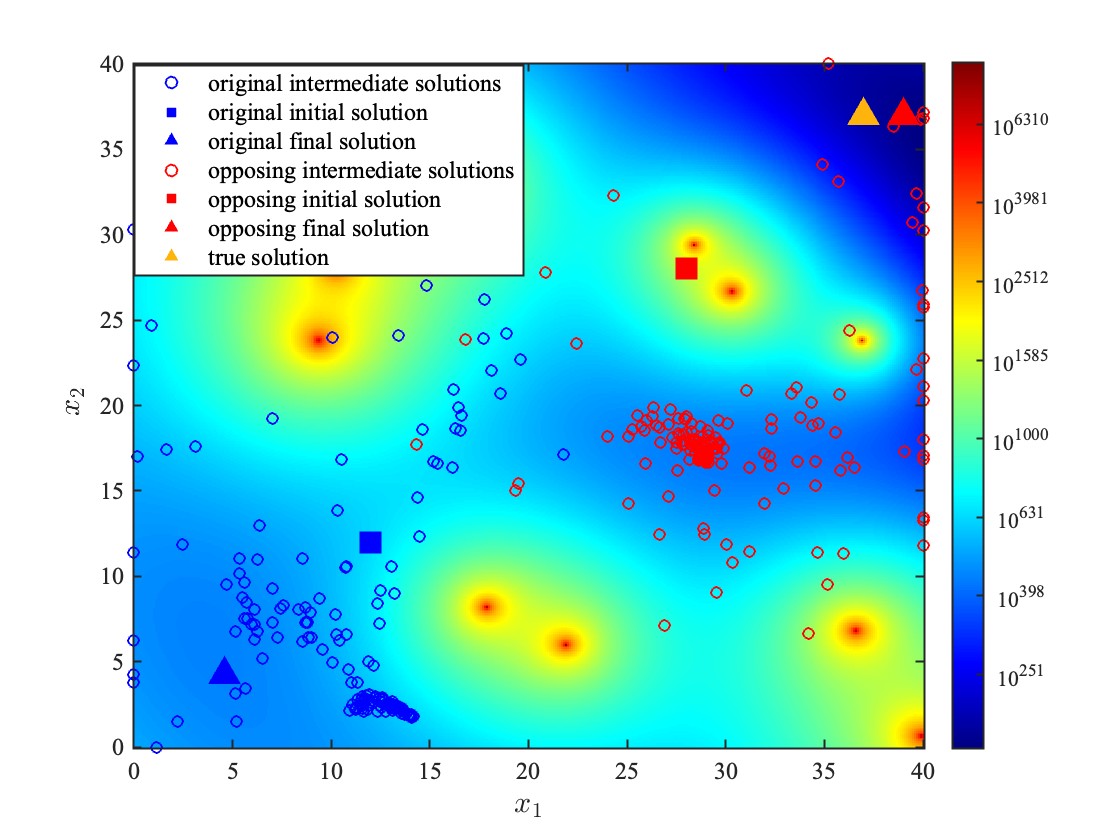}
\caption{Visualization of the proposed algorithm's optimization paths in RSS-based localization.}\label{fig:visualization_opimization_path}
\end{figure}

Figure.\,\ref{fig:visualization_opimization_path} illustrates a specific optimization path scenario in the proposed algorithm for the RSS-based localization problem. The search process starting from the original initial solution might fall into a local optimum. However, due to its substantial distance from the original, the search process beginning from the opposing initial solution avoids the local optimum. This enhances the possibility of reaching the global optimum, underscoring the importance of the OBL in tackling complex optimization problems.

In our SAA-based algorithm, we utilize the concepts of a current intermediate solution, denoted as $\boldsymbol{x}$, and a next intermediate solution, represented as $\boldsymbol{x}_\text{new}$. The next intermediate solution is randomly generated in the vicinity of the current intermediate solution. Depending on certain criteria, the current intermediate solution might either move to the next intermediate solution or remain unchanged. The decision to move is governed by the value of the ML cost function, which is defined as
\begin{align}
f(\boldsymbol{x})=\sum\limits_{i=1}^{N}{{\left ({P_i}-{P_0}+10\gamma {{\log }_{10}}\frac{\left\| \boldsymbol{x}-{\boldsymbol{a}_i} \right\|}{d_0}\right )}^2},
\end{align}
where we assume the standard deviations of $n_i$ are equal, i.e., $\sigma_i=\sigma$, for simplicity. 

If the ML cost value of the next intermediate solution is less than that of the current one, the movement is always accepted. However, when the ML cost value of the next intermediate solution is larger, the decision to accept or reject the move is determined by the Metropolis criterion as
\begin{align}
p_{accept}=
\left\{
	\begin{aligned}
	\text{e}^{-\frac{f(\boldsymbol{x}_{new})-f(\boldsymbol{x})}{kT}}, \quad f(\boldsymbol{x})<f(\boldsymbol{x}_{new})\\
	1, \quad \quad  \quad \quad f(\boldsymbol{x}) \ge f(\boldsymbol{x}_{new})\\
	\end{aligned}
	\right.,
\end{align}
where $T$ denotes to  the absolute temperature, $k$ denotes to the Boltzmann constant.

\begin{figure}
\centering
\includegraphics[width=1\linewidth]{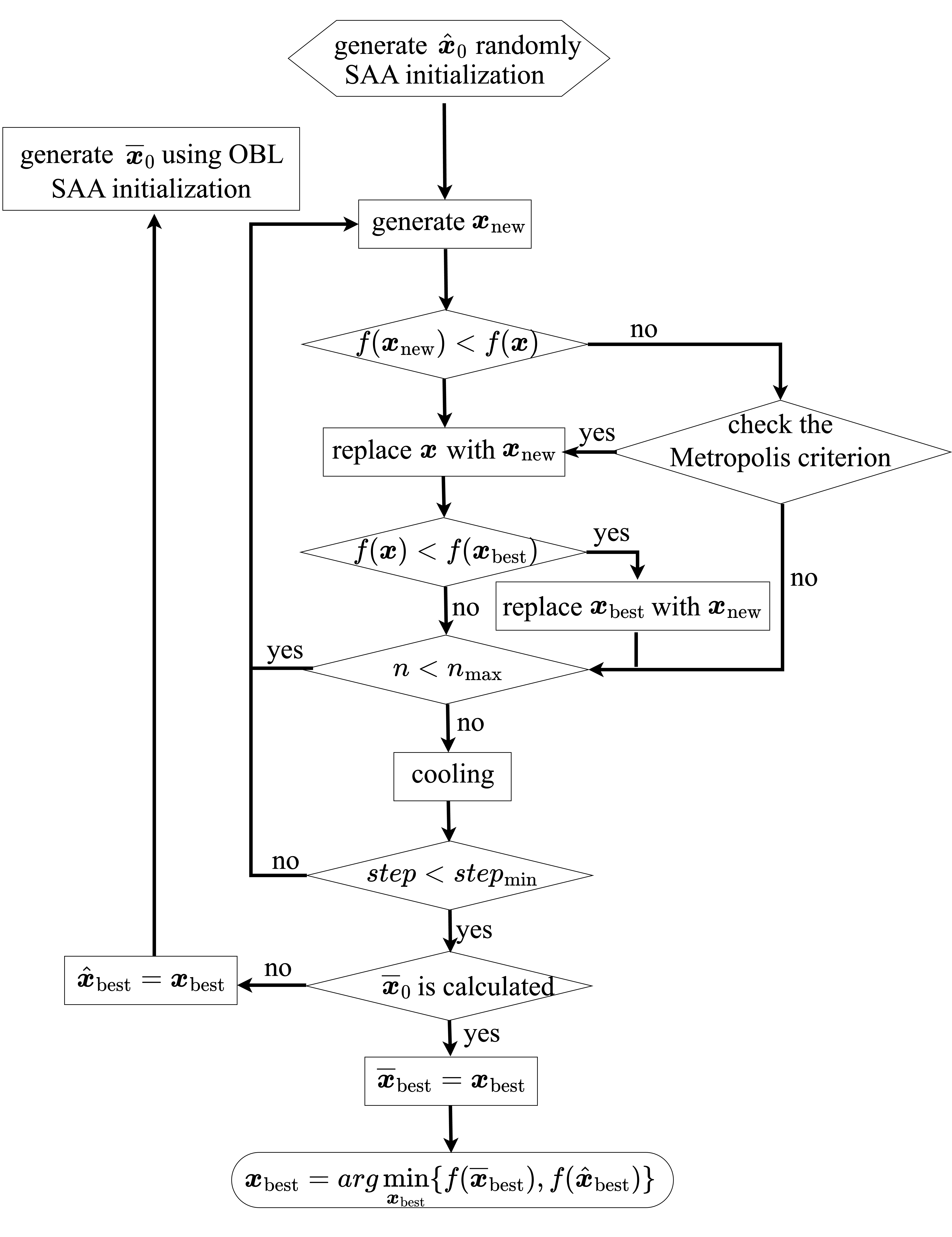}
\caption{Flowchart of the proposed algorithm.}\label{flowchart}
\end{figure}

The simulated annealing algorithm computes the two optimal values from the two initial solutions and selects the one with the smaller cost function as the target node’s location. Fig.\,\ref{flowchart} summarizes this section with a flowchart. The SAA has some control parameters that need to be specified, such as $\lambda$, which represents the step ratio at each cooling, $\epsilon$, which represents the temperature ratio at each cooling, and $n_{max}$, which stands for the maximum number of iterations. The next section determines these hyperparameters through simulation results.

\section{Simulation Results}
This section presents various simulation results to determine the optimal settings of the proposed algorithm and evaluate its performance. We use Monte-Carlo simulations with 10000 samples in a 40\,m$\times$40\,m area to obtain all the results. We generate the measurements between the anchor nodes and the target node using (\ref{e1}), with $P_0=10$\,dB and $\gamma=3$. We specify the values of $N$ and $\sigma$ in each simulation. We mainly use the root mean squared error (RMSE) to measure the localization accuracy of the methods. The RMSE is defined as $\text{RMSE} = \sqrt{\frac{1}{M_c}{{\sum\limits_{i=1}^{M_c}{\left\| {\boldsymbol{x}_{i}}-{{\hat{\boldsymbol{x}}}_{i}} \right\|}^{2}}}}$, where ${\boldsymbol{x}_{i}}$ and ${{\hat{\boldsymbol{x}}}_{i}}$ are the $i$-th target node’s actual location and estimated location, respectively, and $M_c$ is the number of Monte-Carlo iterations.

\subsection{Control Parameters in the SAA}
The SSA has a set of parameters that significantly affect its performance. We need to select and control these parameters carefully during the algorithm execution to ensure that the SSA finds the optimal solution with the best search path. We use three computer simulations to determine the optimal values of these parameters for the proposed algorithm.

 Table\,\ref{table:dT} shows the RMSE and the average running time (RT) for different values of $\epsilon$, with $\lambda =0.4$, $n_{max}=500$, $N=10$, $\sigma=2$\,dB. The table indicates that the average RT decreases as $\epsilon$ increases, while the RMSE remains stable for different values of $\epsilon$. Hence, we choose $\epsilon=0.9$ as a reasonable value. 
 
 Table\,\ref{table:dstep} compares the RMSE and the average RT for different values of $\lambda$, with  $\epsilon=0.9$, $n_{max}=500$, $N=10$, $\sigma=2$\,dB. The table reveals that the RMSE slightly improves, but the average RT increases significantly when $\lambda>0.4$. Therefore, we select $0.4$ as a reasonable value for $\lambda$. 
 
 Table\,\ref{table:n_max} displays the RMSE and the average RT for different values of $n_{max}$, with $\epsilon=0.9$, $\lambda=0.4$, $N=10$, $\sigma=2$\,dB. The table demonstrates that the RMSE does not change when ${n_{max}\geq 500}$. Consequently, we consider $n_{max}=500$ as a reasonable value.

\begin{table}[!t]
\caption{The RMSE and the average running time versus $\epsilon$.}\label{table:dT}
\centering
\begin{tabular}{c c c c c c c}\toprule
\hline
$\epsilon$\ & 0.2 & 0.4 & 0.6 & 0.8 & 0.9 & 0.95\\
\hline
RMSE (m) & 1.85 & 1.85 & 1.85 & 1.85 & 1.85 & 1.86\\

RT (s) & 0.0067 &  0.0059 & 0.0057  & 0.0057 & 0.0056 & 0.0056\\
\bottomrule
\end{tabular}
\end{table}

\begin{table}[!t]
\caption{The RMSE and the average running time versus $\lambda$.}\label{table:dstep}
\centering
\begin{tabular}{c c c c c c c}\toprule
\hline
$\lambda$ & 0.2 & 0.3 & 0.4 & 0.5 & 0.6 & 0.8\\
\hline
RMSE (m) & 1.91 & 1.88 & 1.87 & 1.87 & 1.87 & 1.87\\

RT (s) & 0.0037 &  0.0046 & 0.0056  & 0.0065 & 0.0084 & 0.0085\\
\bottomrule
\end{tabular}
\end{table}

\begin{table}[!t]
\caption{The RMSE and the average running time versus $n_{max}$.}\label{table:n_max}
\centering
\begin{tabular}{c c c c c c c}\toprule
\hline
$n_{max}$ & 200 & 300 & 400 & 500 & 600 & 800 \\
\hline
RMSE (m) & 2.15 & 1.90 & 1.85 & 1.84 & 1.84 & 1.84\\

RT (s) & 0.0023 &  0.0034 & 0.0046  & 0.0057 & 0.0068 & 0.0090\\
\bottomrule
\end{tabular}
\end{table}

\subsection{Comparison with the existing algorithms}
We compare the proposed algorithm with four other localization algorithms to demonstrate its high localization accuracy. The four algorithms are the DEOR in \cite{DEOR}, the DEOR-fast in \cite{DEORfast}, the LSRE-SDP in \cite{SR-SDP}, and the SR-WLS in \cite{GTRS}. We also present the Cramer-Rao lower bounds (CRLB) as the theoretical lower bound for the performance of any unbiased localization algorithms.

\begin{figure}
\centering
\includegraphics[width=1\linewidth]{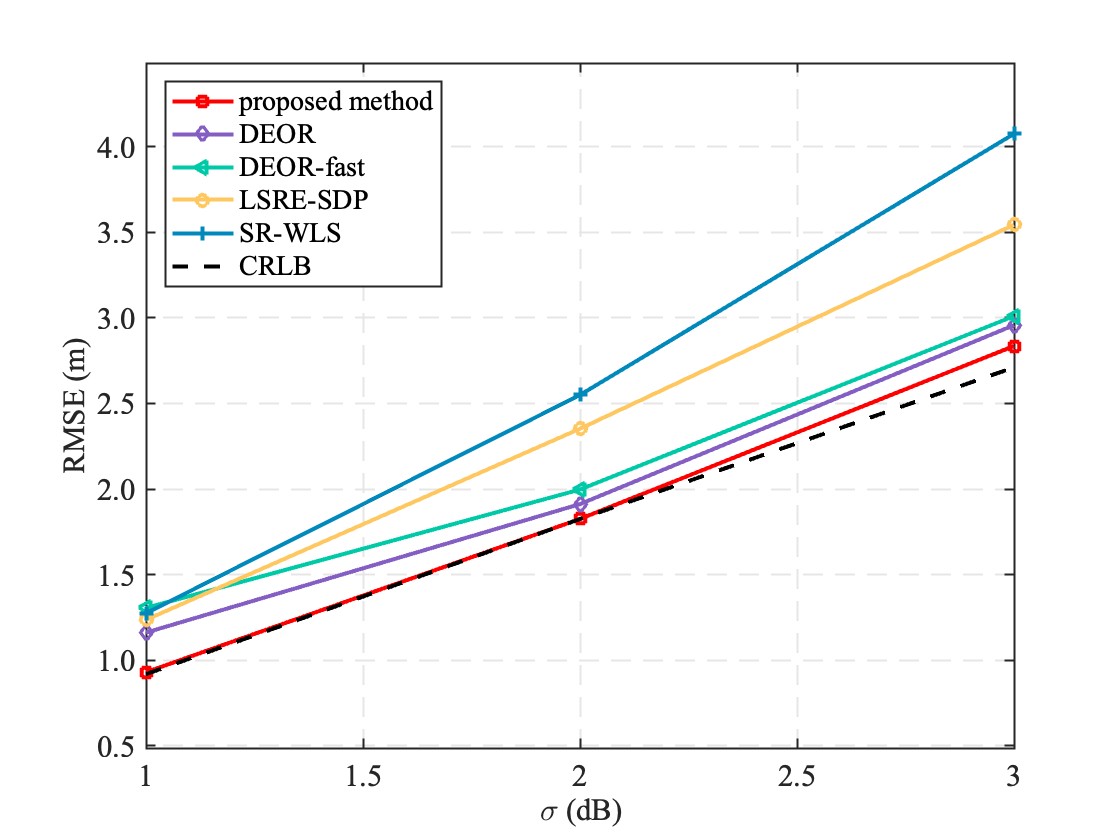}
\caption{RMSE versus $\sigma$ comparison, when $N=10$.}\label{f2}
\end{figure}

Fig.\,\ref{f2} plots the RMSE against $\sigma$, with $N=10$. The figure shows that the RMSE of all the algorithms, including the CRLB, increases as $\sigma$ increases. This is expected because a larger measurement noise results in a larger localization error. The proposed algorithm has a lower RMSE than the four compared algorithms, and it is very close to the CRLB. This indicates the high localization accuracy of the proposed algorithm.

\begin{figure}
\centering
\includegraphics[width=1\linewidth]{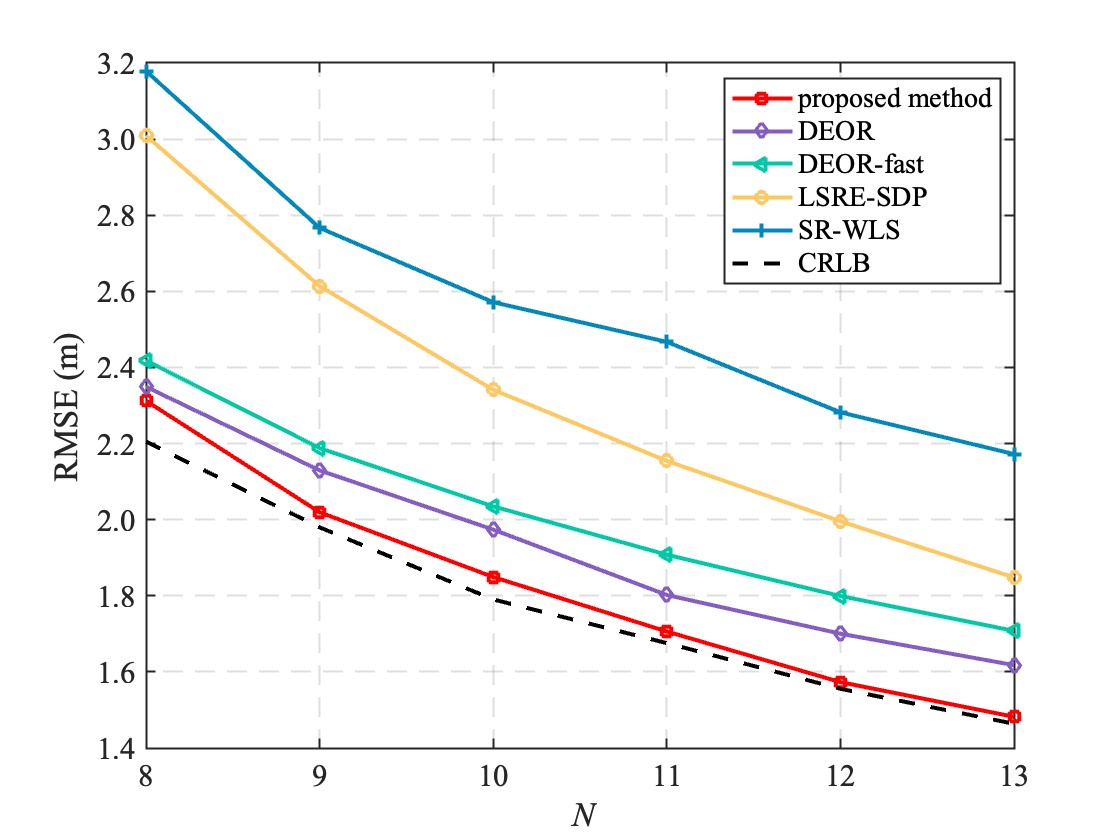}
\caption{RMSE versus $N$ comparison, when $\sigma=2$\,dB.}\label{f3}
\end{figure}

Fig.\,\ref{f3} illustrates the RMSE for different values of $N$, with $\sigma=2$\,dB. The RMSE of the considered algorithm improves as $N$ increases. This is because more measurements improve the localization accuracy. The proposed algorithm has a lower RMSE than the four compared algorithms and it is very close to the CRLB. This confirms the high positioning accuracy of the proposed algorithm.

\section{Conclusion}
Accurate localization is a crucial technique in wireless sensor networks. This letter presents a novel localization algorithm based on the RSS measurements. The algorithm uses the SAA and the OBL to simplify the procedure and enhance localization accuracy. The algorithm determines the original initial solution randomly and uses OBL to generate the opposing initial solution. Then it finds the optimal solution for the target location estimation by SAA. The simulation results show that the proposed algorithm achieves higher localization accuracy than the compared algorithms.

\end{document}